\documentclass[11pt]{article}
\usepackage{graphicx}
\def\title{\begin{center}\Large\bf}
\def\author(s){\vspace{0.3cm}\large\rm}
\def\text{\end{center}}

\begin{document}


\noindent {\small{\it Submitted to the Proceedings of the
International Conference "Near-Earth Astronomy - 2007", 3-7
September 2007, Terskol settlement, Kabardino-Balkaria, Russia}}

\medskip \hrule \medskip

\bigskip
\bigskip


\title
Detection of an oscillatory phenomenon in optical transient
counterpart of GRB090522C from observations on Peak Terskol

\bigskip



\author(s)
B.E. Zhilyaev $^1$, M.V. Andreev $^2$, A.V. Sergeev $^2$, V.B.
Petkov $^3$

\bigskip

\smallskip

\noindent $^1$ {\small {\it Main Astronomical Observatory, NAS
of Ukraine,  27 Zabolotnoho, 03680 Kiev, Ukraine}} \\

\noindent {\small {\it e-mail:}} {\small{\bf zhilyaev@mao.kiev.ua}}

\smallskip

\noindent $^2${\small {\it International Centre for Astronomical,
Medical and Ecological Research \\ Terskol settlement,
Kabardino-Balkaria, 361605 Russia}} \\

\smallskip

\noindent $^3${\small {\it Baksan Neutrino Observatory, INR RAS, Neutrino, Russia}} \\


\text


\section*{Abstract}

22 Sep 2005 Swift-BAT triggered and located GRB050922C. The light
curve shows the intense broad peak with $T_{90}$ of $(5 \pm 1)$ s.
The Nordic Optical Telescope has obtained spectra of the afterglow
with several absorption features corresponding to a redshift of $z =
2.17 \pm 0.03$. Observation of optical transient of GRB050922C was
carried out in the R-band with the 60-cm telescope equipped with a
CCD on Peak Terskol (North Caucasus). The OT magnitude was fading
from R $\approx 16$ to $\approx 17.5$. Detection of an oscillatory
phenomenon in the R post-burst light curve is described in this
work. Analysis of the R data reveals coherent harmonic with a period
of $0.0050 \pm 0.0003$ days (7.2 min) during observing run of about
0.05 days ($\sim 70$ min). Amplitude of oscillations is about 0.05
magnitude. The simplest model suggests that GRB050922C may result
from tidal disruption of a white dwarf star by a black hole of about
one thousand solar mass. The periodicity in the light curve can be
identified with relativistic precession of an accretion disc.

\vspace{1.0cm}

\noindent {{\bf keywords}$\,\,\,\,$\large gamma-rays: bursts --
gamma-rays: theory -- methods: statistical -- black hole physics }

\large

\section{Introduction}

Gamma-ray burst GRB050922C was detected with satellites (Swift,
HETE) and ground-based instruments. 22 Sep 2005 at 19:55:50 UT
Swift-BAT triggered and located GRB050922C (Trigger = 156467). The
light curve in Fig 1 shows the intense broad peak starting from T -
3 to T + 3 s with two sub-peaks on top. Ground-based instruments
were able to measure an afterglow during a few hours both
photometrically and spectrally. The Nordic Optical Telescope (La
Palma) has obtained spectra of the afterglow [1]. It found several
absorption features, including strong Lyman-alpha, OI+SiII, CII,
SiIV, CIV, AlII and AlIII, corresponding to a redshift of $z = 2.17
\pm 0.03$. Assuming z = 2.17 and a standard cosmology model the
isotropic energy release is $E_{iso} \sim 8\cdot 10^{52}$ erg, the
maximum luminosity is $(L_{iso})_{max} \sim 1.6\cdot 10^{53}$ erg/s.


\section{Observation of optical transient of GRB050922C }
\noindent The ROTSE-IIID instrument at Turkey started ground-based
observations of GRB 050922C afterglow at $T_{burst} + 172.4$ s
($T_{GCN}=6.8$ s). It interrupted observations at 651 s due to
precipitation. Observation of optical transient of GRB050922C were
continued in the R-band with the 60-cm telescope equipped with a CCD
camera on Peak Terskol (North Caucasus) starting Sep 22, 2005,
20:08:45 UT [2]. 61 images of 60 s exposure were taken between
20:08:45 and 21:18:20 UT. The OT magnitude was fading from R
$\approx$ 16 to $\approx$ 17.5. Detection of an oscillatory
phenomenon in the R post-burst light curve is described in this
work.

\begin{figure*}
\centering
\resizebox{0.90\hsize}{!}{\includegraphics[angle=000]{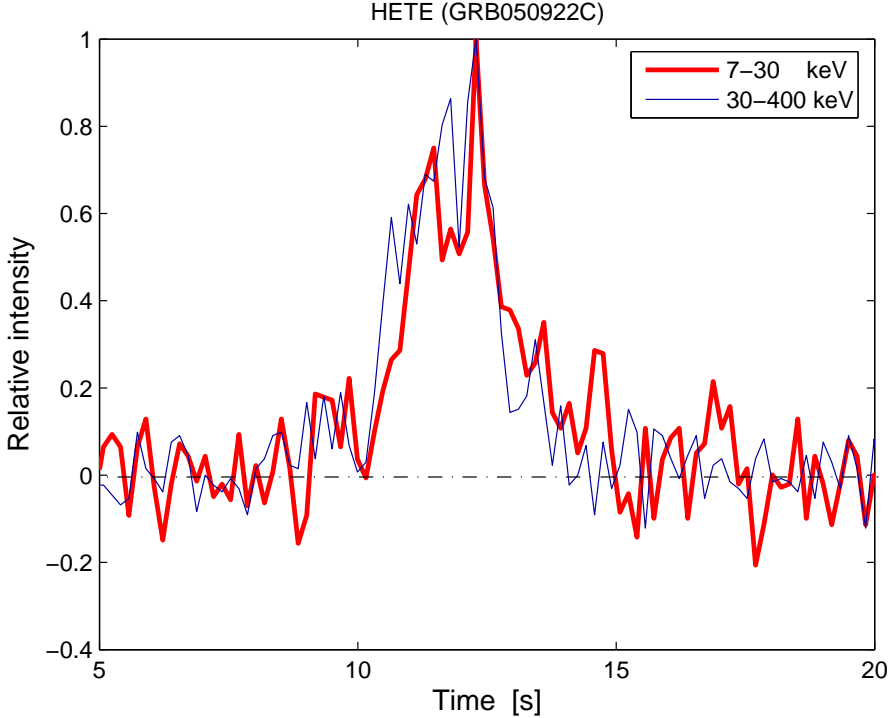}}
\caption{The GRB050922C light curve from HETE [3].}
\end{figure*}

\section{Results}

\noindent The original and resample equally spaced data are shown in
Fig 2. We have used piecewise cubic spline interpolation to fill up
a few gaps inside the primary data. High-frequency residuals after
removing of polynomial trend are shown in Fig 3. The spectral
analysis of residuals with the Tukey spectral window reveals clearly
a harmonic with a period of 0.005 d = 7. 2 min (Fig 4). The same
result with the harmonic period of 0.0050 d demonstrates the wavelet
power spectra in Fig 5. To make clear coherence features of this
harmonic the reconstruction of a signal was performed using a
continuous wavelet transform [5]. Digital filtering with the help of
a continuous wavelet transform allows one to separate the
non-stationary signal in an appropriate frequency range. This
permits the oscillation to be analyzed separately.  These results
are shown in Fig 5. We may conclude that the GRB050922C light curve
in the R band reveals coherent harmonic with a period of 0.0050 d
(7.2 min) during observing run of about 0.05 d ($ \sim 70$ min). In
Fig 7 a harmonic period of $0.0050 \pm 0.0003$ d follows from the
maximum coordinates measurements shown in Fig 6. Amplitude of
oscillations is about 0.05 mag and comparable with the internal
accuracy of the photometry. Note both the windowed and wavelet power
spectra show coherent oscillation at the confidence level more than
99\%.

\begin{figure*}
\centering
\resizebox{0.90\hsize}{!}{\includegraphics[angle=000]{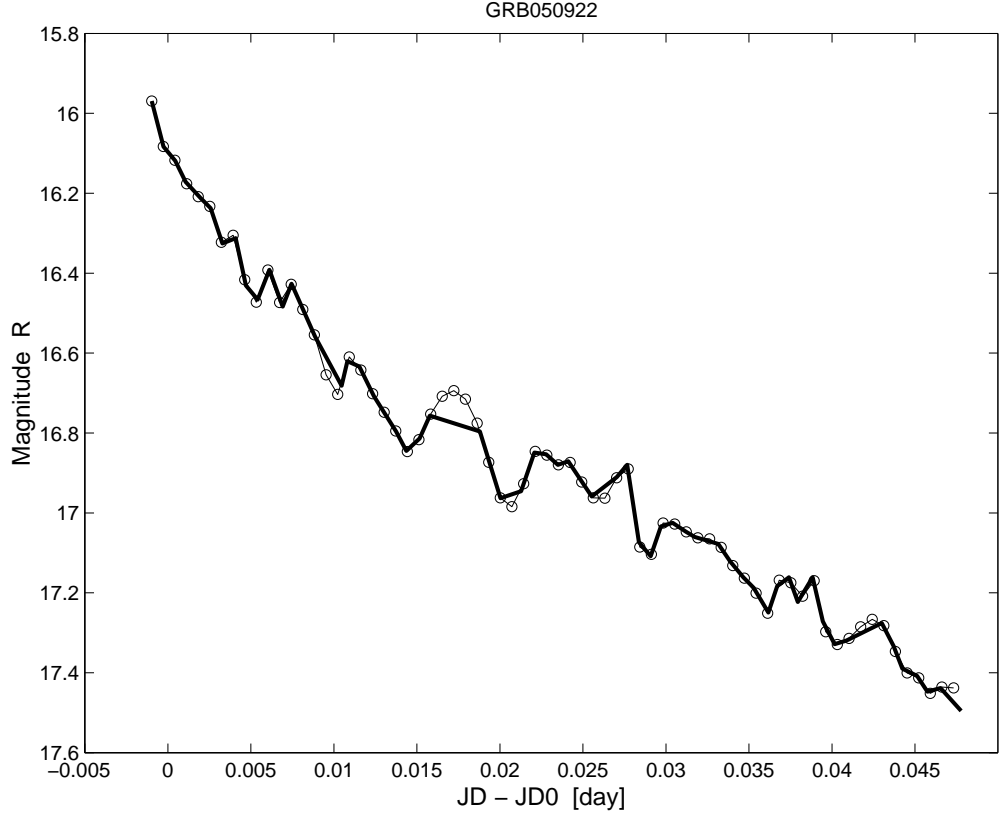}}
\caption{The original (the heavy curve) and resample equally spaced
data (circles) to fill up a few gaps inside the primary data are
shown.}
\end{figure*}

\begin{figure*}
\centering
\resizebox{0.90\hsize}{!}{\includegraphics[angle=000]{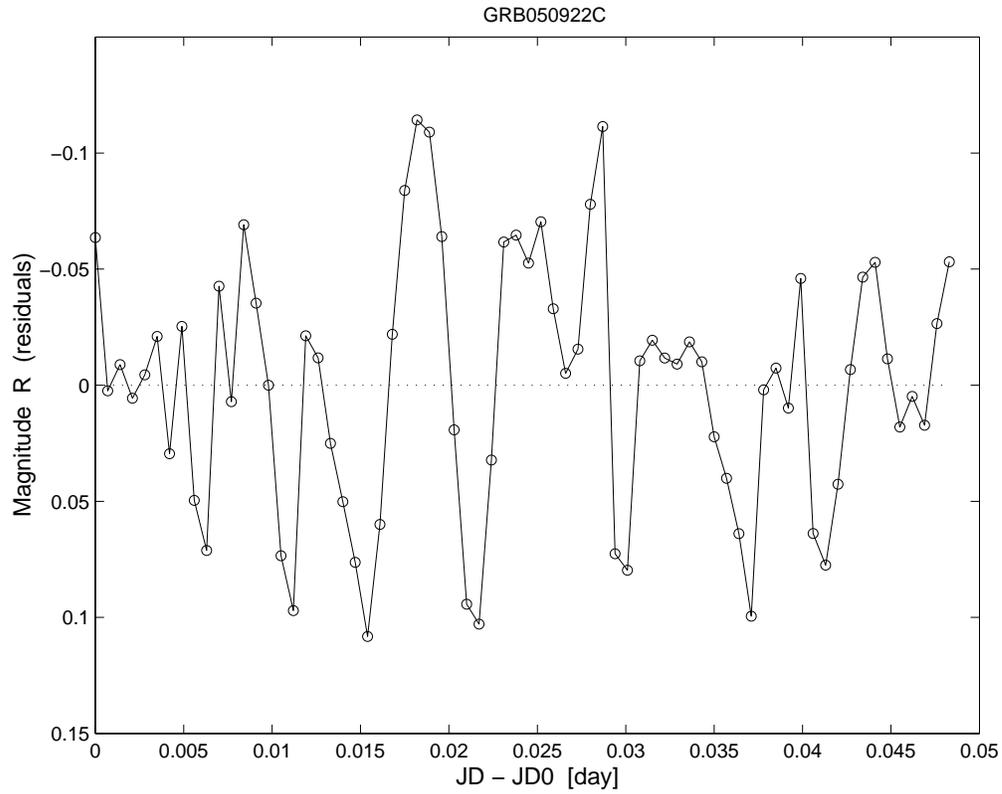}}
\caption{High-frequency residuals after removing of polynomial
trend. }
\end{figure*}

\begin{figure*}
\centering
\resizebox{0.90\hsize}{!}{\includegraphics[angle=000]{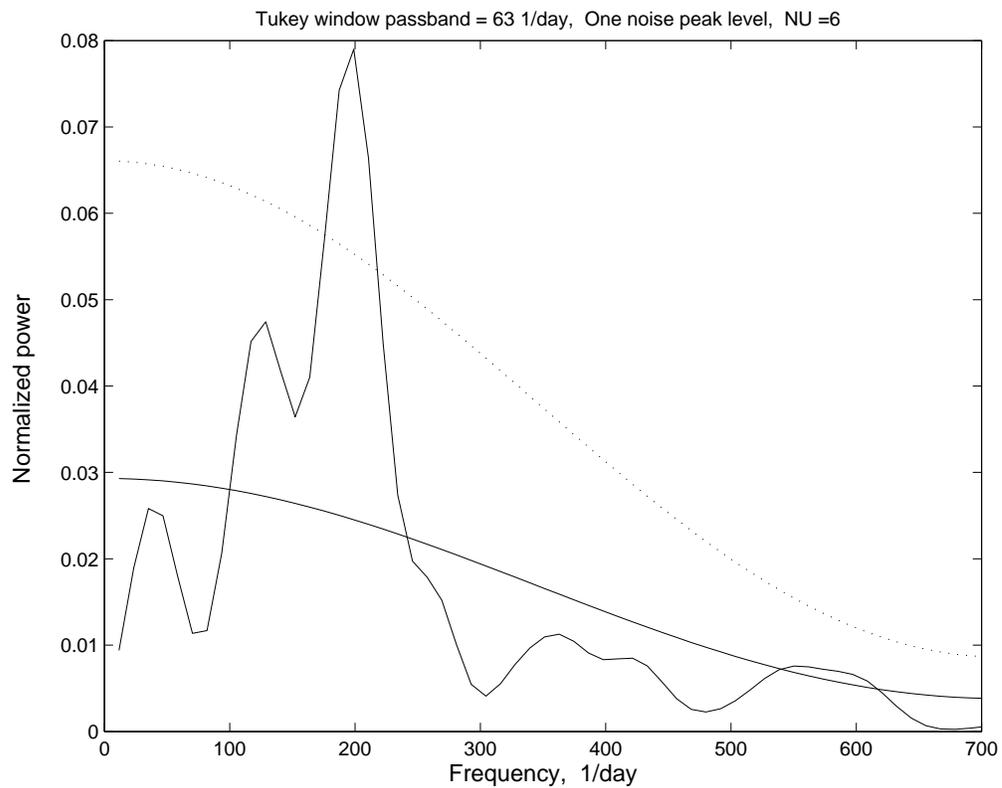}}
\caption{The power spectrum of residuals with the Tukey spectral
window reveals clearly a harmonic with a period of 0.005 d = 7. 2
min. The doted line corresponds to the 99\% confidence for noise
peaks.}
\end{figure*}

\section{Discussion}

\noindent A luminous gamma/X-ray burst can occur when a star passes
within the tidal radius of the massive black hole, and is disrupted.
Disruption begins when the tidal acceleration due to the black hole
equals to the self-gravity of the star.  Light curve of the stellar
debris is dependent both on the mass and spin of the black hole and
disrupted star. The tidal disruption time scale is about of the
free-fall time $T_{ff}$. This is the characteristic time it would
take a body to collapse under its own gravitational attraction, if
no other forces existed to oppose the collapse
\begin{equation}\label{free}
    T_{ff} = \frac{1}{4}\sqrt{\frac{3\pi}{2G\rho}}
\end{equation}
Here $\rho$ is the mean density. Note also that the free-fall
timescale is practically coinciding with an oscillation period of a
self-gravitating body
\begin{equation}\label{osscil}
    P\approx \sqrt{\frac{4\pi}{G\rho}}
\end{equation}
Numerical calculation of the tidal disruption gives the estimate of
the dynamical time of disruption close to above mentioned values,
too [4]. For the Sun $T_{ff}= 1.78\cdot 10^{3}$ s = 29.7 min; for a
red dwarf of M5 V class $T_{ff} = 685$ s = 11.8 min; for a white
dwarf and a neutron star of the solar mass $T_{ff} = 1.78$ s and
$\approx 0.0001 = 0.1 $ ms, respectively. Really we can suppose that
GRB050922C, bearing in the mind the above estimate of $T_{ff}$,
results from tidal disruption of a white dwarf star by a black hole.
Postdisruption behavior concerns formation of an accretion torus
around the BH.
\begin{figure*}
\centering
\resizebox{0.90\hsize}{!}{\includegraphics[angle=000]{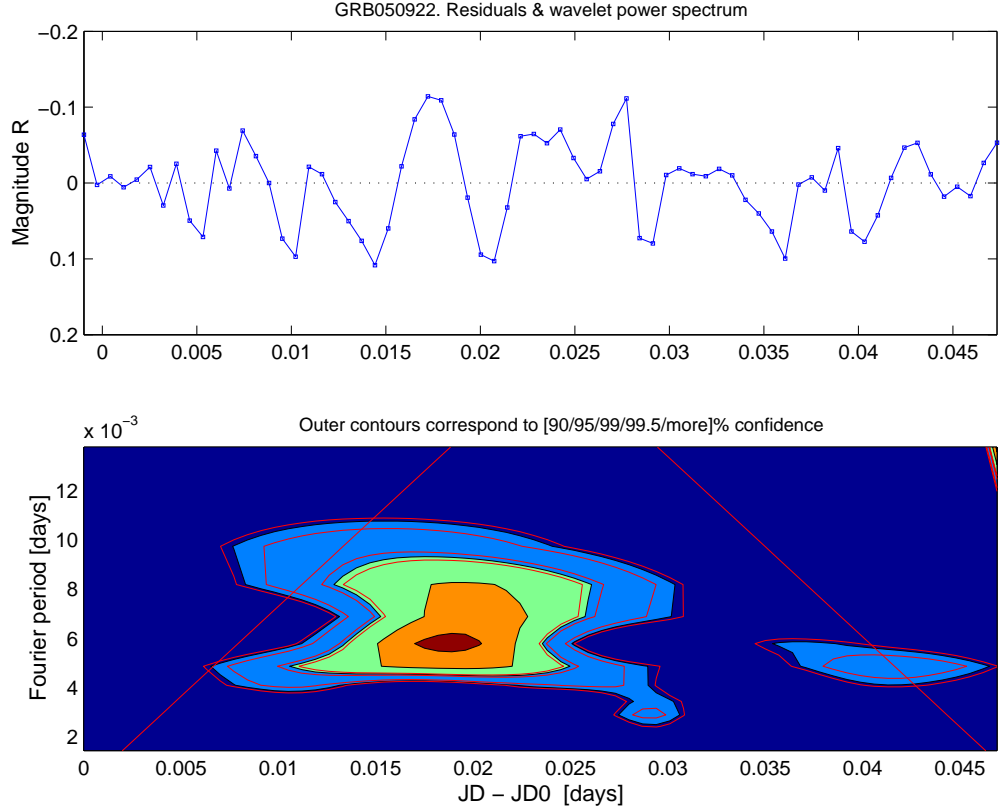}}
\caption{High-frequency residuals (top panel) and their wavelet
power spectrum (bottom panel). The outer contours correspond to 90\%
and more confidence . A harmonic period is of 0.0050 d. The solid
lines represent the cone of influence in the wavelet diagram where
edge effects become important. We apply the wavelet transform
following the approach laid out by Torrence \& Compo [5].}
\end{figure*}

\section{The tidal disruption of a white dwarf by a massive black hole}
\noindent The pericentric distance at disruption is [4]
\begin{equation}\label{disradius}
   R_{p}\approx R_{star}(M_{BH}/M_{star})^{1/3}
\end{equation}
The tidal disruption radius for a white dwarf (WD) of the solar mass
and radius of 0.01 $R_{\odot}$ depending on the black hole mass is
shown in Fig 8.  The pericentric distance is in the $r_{g}$ units,
where $r_{g} = 2GM_{BH}/c^{2}$ - the gravitational radius of a black
hole. If the BH mass is $\geq  10^{38}$ g WD plunges in BH without
disruption. So we may assert that in our case the BH mass is no more
than $10^{5}$ solar mass. From the tidal disruption condition (3)
for the Keplerian period of disrupted star we have
\begin{equation}\label{kepler1}
    P^{2}=4\pi/GM_{BH}R^{3}_{p}
\end{equation}
From Eq (3) for $R_{p}$ after some algebra we have also
\begin{equation}\label{kepler2}
    P^{2}=4\pi/GM_{star}R^{3}_{star}
\end{equation}

\begin{figure*}
\centering
\resizebox{0.90\hsize}{!}{\includegraphics[angle=000]{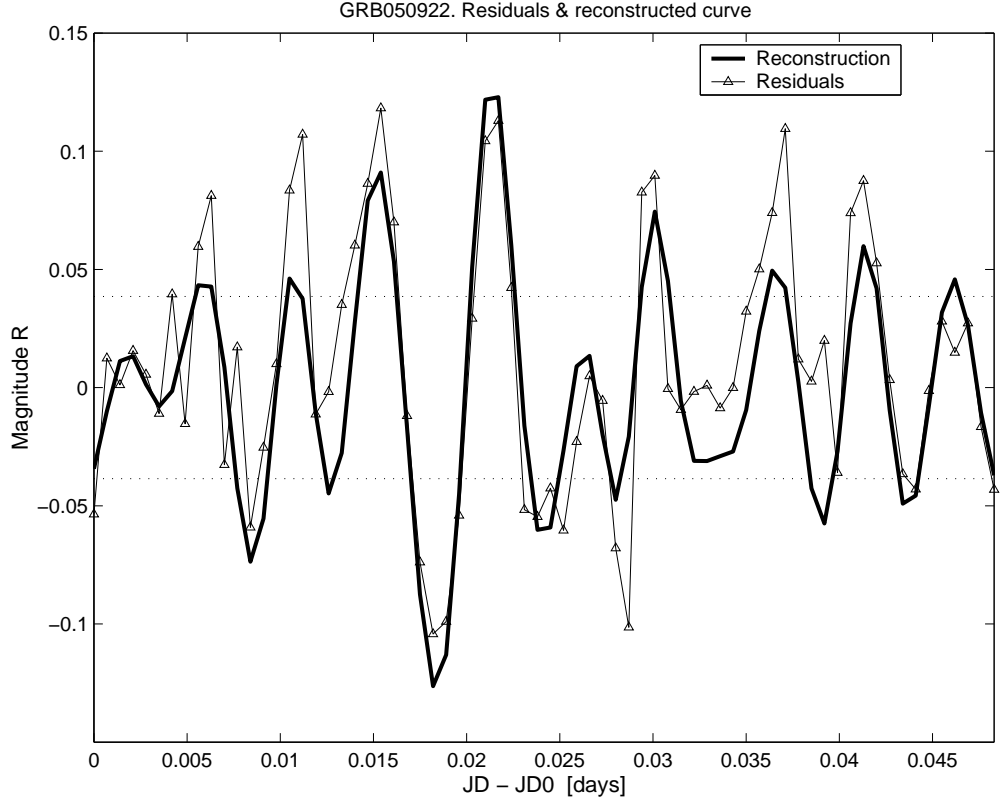}}
\caption{The reconstructed signal (the heavy curve) and raw data.
The 2-sigma error corridor is shown as the dotted lines.}
\end{figure*}

Thus, Keplerian period at the moment of disruption is independent of
the BH mass. It depends only on the mass and radius of the star
itself. We can sum up the main findings as follows:
\begin{itemize}
  \item The burst light curve showed several peaks spaced by
  7.2 min, resembling about 5\% periodic modulation of the overall light
  curve profile.
  \item The periodicity in the light curve can be identified either
  with Kepler orbital motion or with relativistic precession of an
  accretion disc.
  \item The Keplerian period at the moment of disruption is too small
  ($P \approx 10$ sec) to be useful.  Surprisingly, it does not depend on
  the BH mass.
\end{itemize}

\begin{figure*}
\centering
\resizebox{0.90\hsize}{!}{\includegraphics[angle=000]{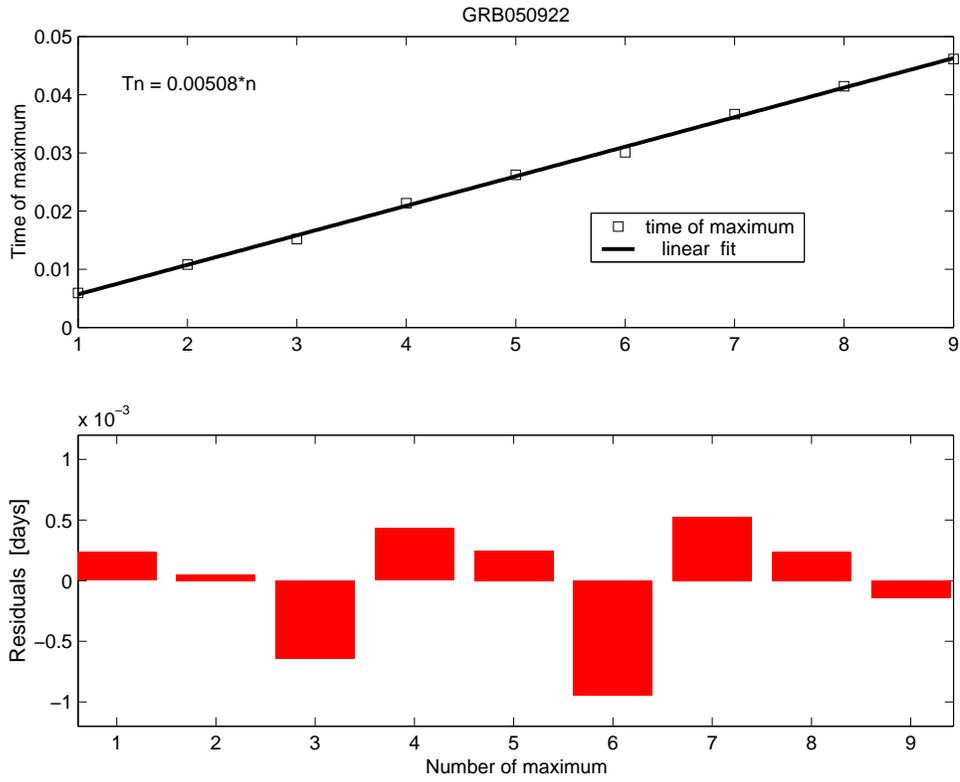}}
\caption{The harmonic period of $0.0050 \pm 0.0003$ d follows from
the linear fit of the maximum coordinates measurements shown in Fig
6.}
\end{figure*}

\begin{figure*}
\centering
\resizebox{0.90\hsize}{!}{\includegraphics[angle=000]{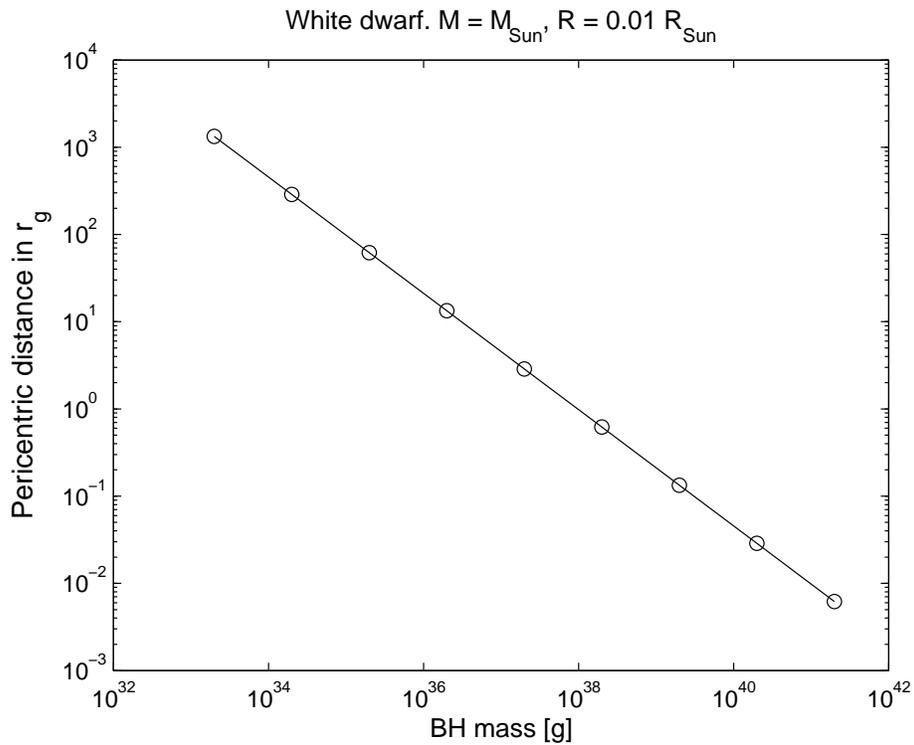}}
\caption{See text}
\end{figure*}

Thus, we can conclude that the periodicity in the light curve can be
identified with    relativistic precession of an accretion disc.
This effect has no analogue in classical mechanics. The precession
in the coordinate system at rest is given by [6]
\begin{equation}\label{precession}
    \Delta \varphi =3\pi GM_{BH}/c^{2}R_{p}
\end{equation}
This equation was lacking to evaluate the BH mass. Why we use
relativistic precession?  Even moderate eccentricity value of the
debris orbit is sufficient to account for modulation of the burst
light curve due to lack of symmetry.

Finally we may evaluate the BH mass of $\sim 10^{3} M_{\odot}$. The
gravitational radius of the BH is about 1800 km. The size of an
accretion disc equal to the tidal disruption radius is 32.6 times
greater.

\section{Conclusion}
\begin{itemize}
  \item GRB050922C may result from tidal disruption of a white dwarf
  star by a massive black hole.
  \item We may conclude that the GRB050922C optical afterglow
  confirms the existence of an intermediate mass black hole, of about
  one thousand solar mass.
  \item The periodicity in the light curve can be identified with
  relativistic precession of an accretion disc.
  \item Both the energy release ($E_{iso} \sim 8\cdot 10^{52}$ erg)
  and the   estimate of the burst event dynamical time scale (some seconds)
  agree   closely with the model of the tidal disruption of a white dwarf star
  by a massive black hole.

\end{itemize}




\begin{thebibliography}{}

\small{ \baselineskip=1pt


\bibitem{Jakobsson06}
{\it P. Jakobsson, J. P. U. Fynbo, C. Ledoux, et al.} 2006, Astron.
Astrophys. 460, L13-L17

\bibitem{Andreev}
{\it Andreev M.V., et al.} GCN 4016

\bibitem{HETE}
{\it } The GRB050922C light curve from HETE:
http://space.mit.edu/HETE/Bursts/GRB050922C/GCN-LCs-U11658.dat

\bibitem{Evans}
{\it C.R. Evans and C.S. Kochanek} 1989, ApJ, 346, L13-L16

\bibitem{Torrence}
{\it Torrence, C. \& Compo, G. P.} 1998, A Practical Guide to
Wavelet Analysis. Bulletin of the American Meteorological Society,
Vol. 79, No 1, 61-78

\bibitem{Misner}
{\it C.W. Misner, K.S.Thorne, J.A. Wheeler} 1973, Gravitation, W.H.
Freeman and Company, San Francisco 1973, Vol. 3

}

\end{thebibliography}
\end{document}